\documentclass[aps,prd,preprint,nopreprintnumbers,nofootinbib,showpacs]{revtex4}

\usepackage{epsfig}
\usepackage{graphicx}
\usepackage{afterpage}
\usepackage{float}
\usepackage{subfigure}
\usepackage{dcolumn}

\def\etal{{\it et al.}}

\begin{document}
\newcommand{\bb}{$b\overline{b}~$}
\newcommand{\qq}{$q\overline{q}~$}
\newcommand{\LLbar}{$\Lambda_b^0\overline{\Lambda}\,\!_b^0~$}
\newcommand{\BBbar}{$B\overline{B}~$}

\preprint{CLNS 04/1895}
\preprint{CLEO 04-14}
\title{Search for $e^+e^- \to$ \LLbar Near Threshold}

\author{D.~Besson}
\affiliation{University of Kansas, Lawrence, Kansas 66045}
\author{T.~K.~Pedlar}
\affiliation{Luther College, Decorah, Iowa 52101}
\author{D.~Cronin-Hennessy}
\author{K.~Y.~Gao}
\author{D.~T.~Gong}
\author{Y.~Kubota}
\author{B.~W.~Lang}
\author{S.~Z.~Li}
\author{R.~Poling}
\author{A.~W.~Scott}
\author{A.~Smith}
\author{C.~J.~Stepaniak}
\affiliation{University of Minnesota, Minneapolis, Minnesota
55455}
% ...
\collaboration{CLEO Collaboration} 
\noaffiliation

\date{Nov. 22, 2004}

\begin{abstract}
Using the CLEO III detector at CESR we study $e^+e^-$ collisions
in the center-of-mass energy close to, or above, \LLbar production
threshold. We search for evidence of \LLbar resonance production
and set upper limits based on inclusive hadron production as a
barometer of \LLbar production.

\end{abstract}
\maketitle

\section{Introduction}
The $\Lambda_b^0$, consisting of $b$, $u$ and $d$ quarks, is the
lowest-lying $b$-flavored baryon, about which comparatively little
is known. Recently the CDF collaboration reported an improved
measurement of the $\Lambda_b^0$ mass \cite{LP2003} of 5620.4
$\pm$ 1.6 $\pm$ 1.2 MeV. The lifetime has long been measured to be
somewhat lower than theoretical expectations \cite{Lblife}. There
is, however,  no measurement available on the direct production of
exclusive \LLbar in $e^+e^-$ annihilation. Such events would be
very useful for establishing absolute branching ratios and other
properties. CLEO has accumulated data using $e^+e^-$ collisions in
the center-of-mass energy range from 11.227 to 11.383 GeV, close
to or just above the \LLbar production threshold. It is possible
to observe a resonant signal, similar to the $\Upsilon (4S)$ for
$B^+$ and $B^0$ mesons, or just an increase in relative production
above threshold. We report here limits on such resonant or
non-resonant production.

\section{Data and Monte Carlo Simulated Sample}
\label{sec:two}
The CLEO III detector is described in detail elsewhere
\cite{CLEOIII_d} \cite{CLEOIII_RICH}. The inner part of the
detector is surrounded by a 1.5 T solenoidal magnetic field. From
the region near the $e^+e^-$ interaction vertex radially outward
it consists of a silicon strip based vertex detector and a drift
chamber used to measure the momenta of charged tracks based on
their curvature. Beyond the drift chamber is a Ring Imaging
Cherenkov Detector, RICH, used to identify charged hadrons,
followed by an Electromagnetic Calorimeter, EC, consisting of
nearly 8000 CsI crystals. Next to the EC there is the solenoidal
coil followed by an iron return path with wire chambers
interspersed in 3 layers to provide muon identification.

This study is based on the total 710 pb$^{-1}$ data sample that
was acquired at 3 MeV intervals between center-of-mass energies,
$E_{CM}$, of 11.227 GeV to 11.383 GeV, to be close to or above
threshold for \LLbar production. The luminosity in each of these
scan points varies from 14 to 20 pb$^{-1}$. In addition, there are
data points taken at a $E_{CM}$ of 11.150 and 11.203 GeV, respectively.
The two data points with lowest and highest energies have
integrated luminosities of 70 and 120 pb$^{-1}$, respectively. We
also use data taken in the four-flavor continuum below the
$\Upsilon (4S)$ to measure the \bb cross section above the
$\Upsilon (4S)$.

For the Monte Carlo, MC, study of the high energy data, we
generated five times more hadronic \qq events than at each beam
energy contained in our data sample. Events were generated
separately for ``light" four-flavor continuum ($ c, s, u, d$) and
\bb continuum events and then combined in the expected 10:1 ratio
absent any resonance production. The decay channels and the
branching fractions of the $\Lambda_b$ are less well known than
the $B^0$ and $B^+$ mesons. We list the $\Lambda_b$ decay modes
and branching fractions we used for the signal Monte Carlo in
Table~\ref{tab:smc}. For the $\Lambda_b^0 \to \Lambda_c^+ \ell^- \bar{\nu}$
branching fraction we re-scaled the $B^0 \to X \ell \bar{\nu}$
branching fraction by the ratio of lifetimes,
$\tau(\Lambda_b)/\tau(B^0)$. The entries denoted by *$q\bar{q}$*
indicate that the processes are generated using a fragmentation
process for the quark-antiquark pair.

\begin{table}[htb]
\begin{center}
\caption{\label{tab:smc} $\Lambda_b$ decay modes and branching fractions
used in the Monte Carlo simulation.}
\begin{tabular}{lccr}
\hline\hline
Decay modes & Branching fraction (\%)\\ \hline
$\Lambda_b \to\Lambda^+_c e^- \nu_e$ & 8.4 \\
$\Lambda_b \to\Lambda^+_c \mu^- \nu_{\mu}$ & 8.4 \\
$\Lambda_b \to\Lambda^+_c \pi^-$ & 4.2 \\
$\Lambda_b \to\Lambda^+_c \rho^-$ & 1.0 \\
$\Lambda_b \to\Lambda^+_c a1^-$ & 2.1 \\
$\Lambda_b \to\Lambda^+_c D^-_s$ & 2.1 \\
$\Lambda_b \to\Lambda^+_c D^{*-}_s $ & 4.2 \\
$\Lambda_b \to\Lambda \eta_c$ & 0.1 \\
$\Lambda_b \to\Lambda J/\psi$ & 0.5 \\
$\Lambda_b \to\Lambda^+_c \pi^-\pi^+\pi^-$ & 2.1 \\
$\Lambda_b \to \Lambda K^0 \pi^-\pi^-\pi^+\pi^-$ & 2.1 \\
$\Lambda_b \to p^+D^0 \pi^-$ & 2.1 \\
$\Lambda_b \to\Lambda^+_c *d\overline{u}* $ & 44.9 \\
$\Lambda_b \to \Sigma^+_c *d\overline{u}* $ & 8.4 \\
$\Lambda_b \to \Omega^+_{cc} *d\overline{u}* $ & 7.3 \\
$\Lambda_b \to p^+ *d\overline{u}* $ & 1.1 \\
$\Lambda_b \to \Xi'^+_c *d\overline{u}* $ & 1.0 \\
\hline\hline 
\end{tabular}
\end{center}
\end{table}

\section{Event Selection}

The major backgrounds to $\Lambda_b$ are non-\bb type hadronic
events, two-photon events ($e^+e^-\to e^+e^- X$) and $\tau^+\tau^-$ pairs. To suppress
these backgrounds we require the following hadronic event
selection criteria:

(1) At least five charged tracks; a track candidate is acceptable if it is a cosine with respect to the beam line of less than 0.9 and
has at least half of the potential tracking chamber hits along its length. This requirement rejects
81\% of the $\tau^+\tau^-$ pairs.

\begin{figure}[hbt]
\epsfig{figure=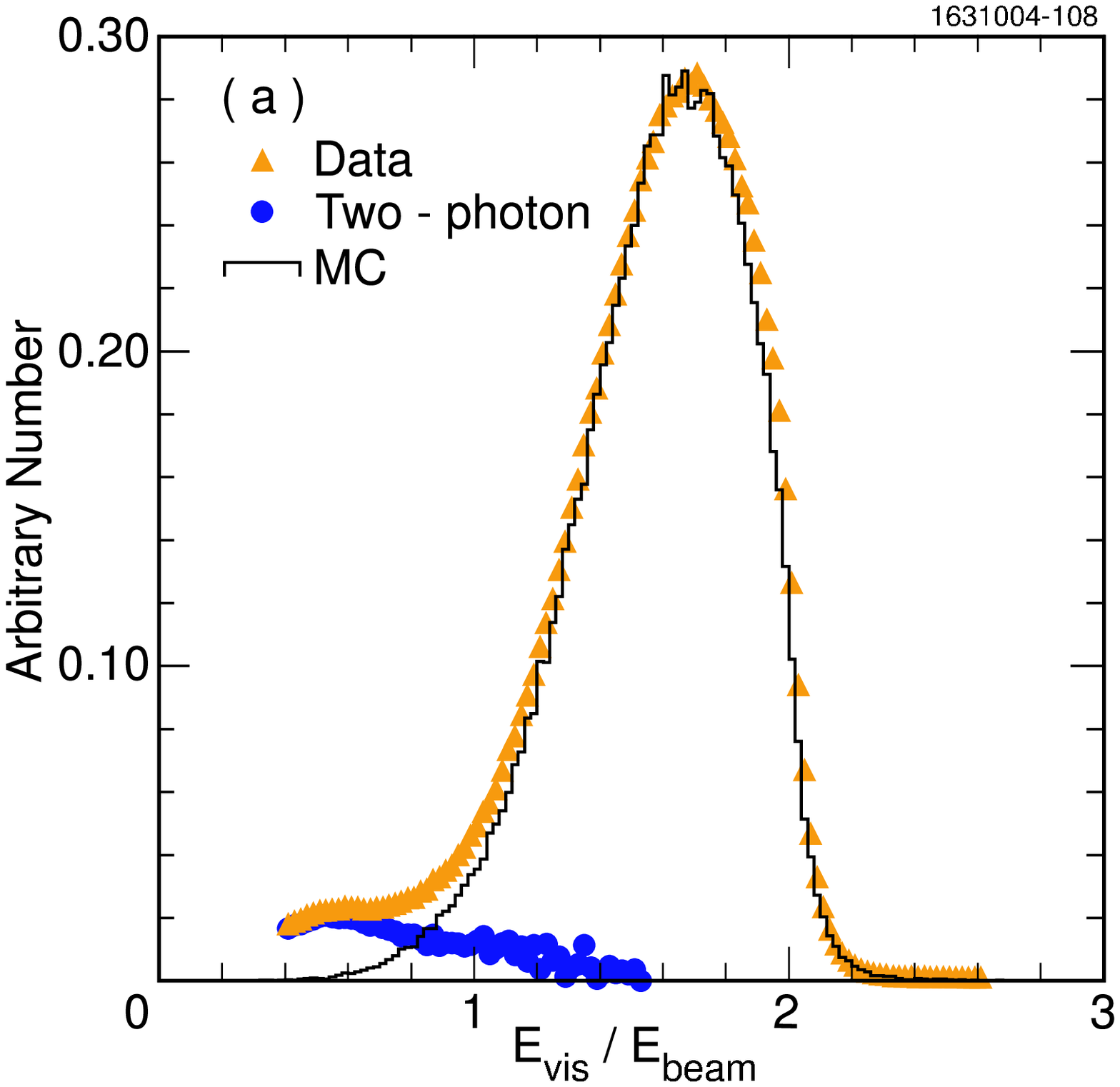,height=3in}
\epsfig{figure=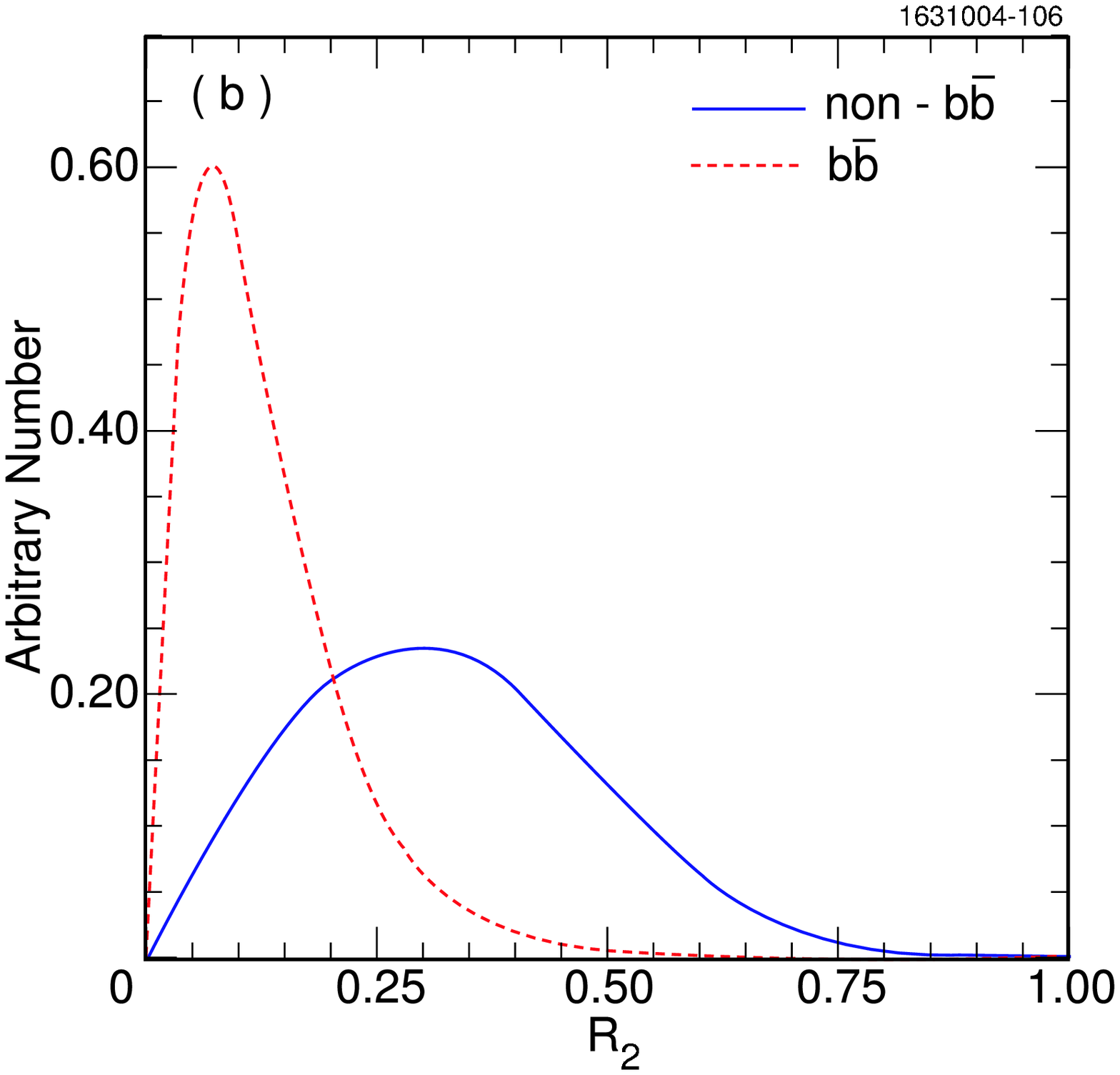,height=3in} \vspace{.1cm} \caption{(a)
$E_{vis}/E_{beam}$ above $\Lambda_b$ threshold data (triangle), five
flavor continuum MC (solid) and simulated two-photon events
(circles). (b) $R_2$ distribution for \bb (dashed) and non-\bb
(solid) type events.} \label{fig:evis2}
\end{figure}

(2) The total visible energy, $E_{vis}$, is required to be greater
than the beam energy, $E_{beam}$.  $E_{vis}$ receives contributions from both charged tracks and unmatched neutral energy clusters greater than 30 MeV. This requirement helps
suppresses two-photon events. Fig.~\ref{fig:evis2}(a) shows the
$E_{vis}/E_{beam}$ distributions  for data, five flavor
Monte-Carlo continuum and simulated two-photon events \cite{two-photon}. Imposing
the requirement $E_{vis}>E_{beam}$ reduces the two-photon
background by 75\% with a small (3\%) loss of hadronic events.

(3) The ratio of the 2$^{nd}$ and 0$^{th}$ Fox-Wolfram moments,
$R_2$, is less than 0.25 \cite{r2paper}. Fig.~\ref{fig:evis2}(b)
shows MC simulated distributions of $R_2$ for both \bb and non-\bb
continuum events. Both areas are normalized to unity. Requiring
$R_2 < 0.25$ selects the more spherically shaped events in
momentum space and greatly enhances the $b\overline{b}$ fraction,
by rejecting 65\% of four-flavor continuum events while losing
only 8\% of the $b\overline{b}$ events.

\begin{figure}[hbt]
\epsfig{figure=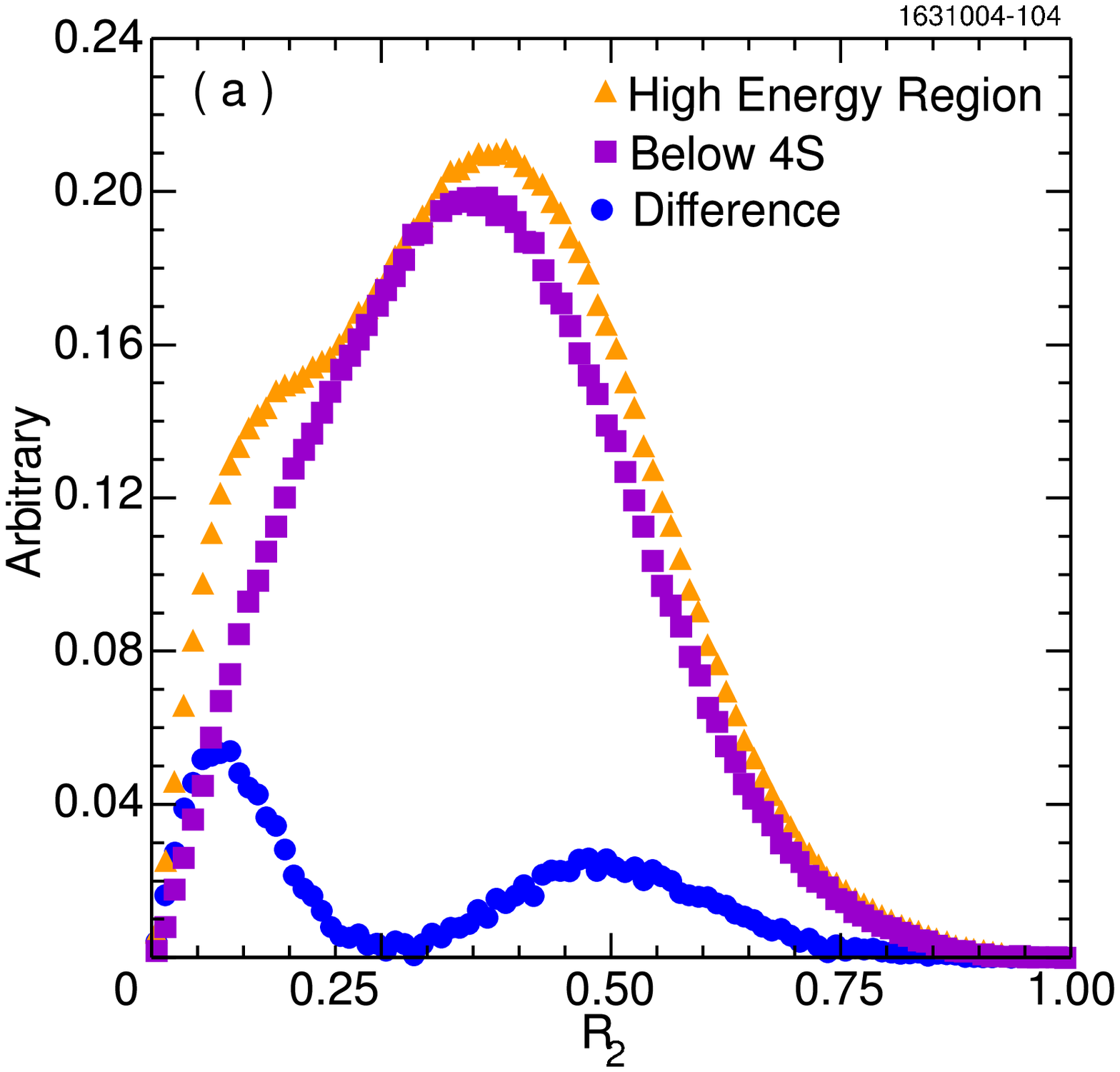,height=3in}
\epsfig{figure=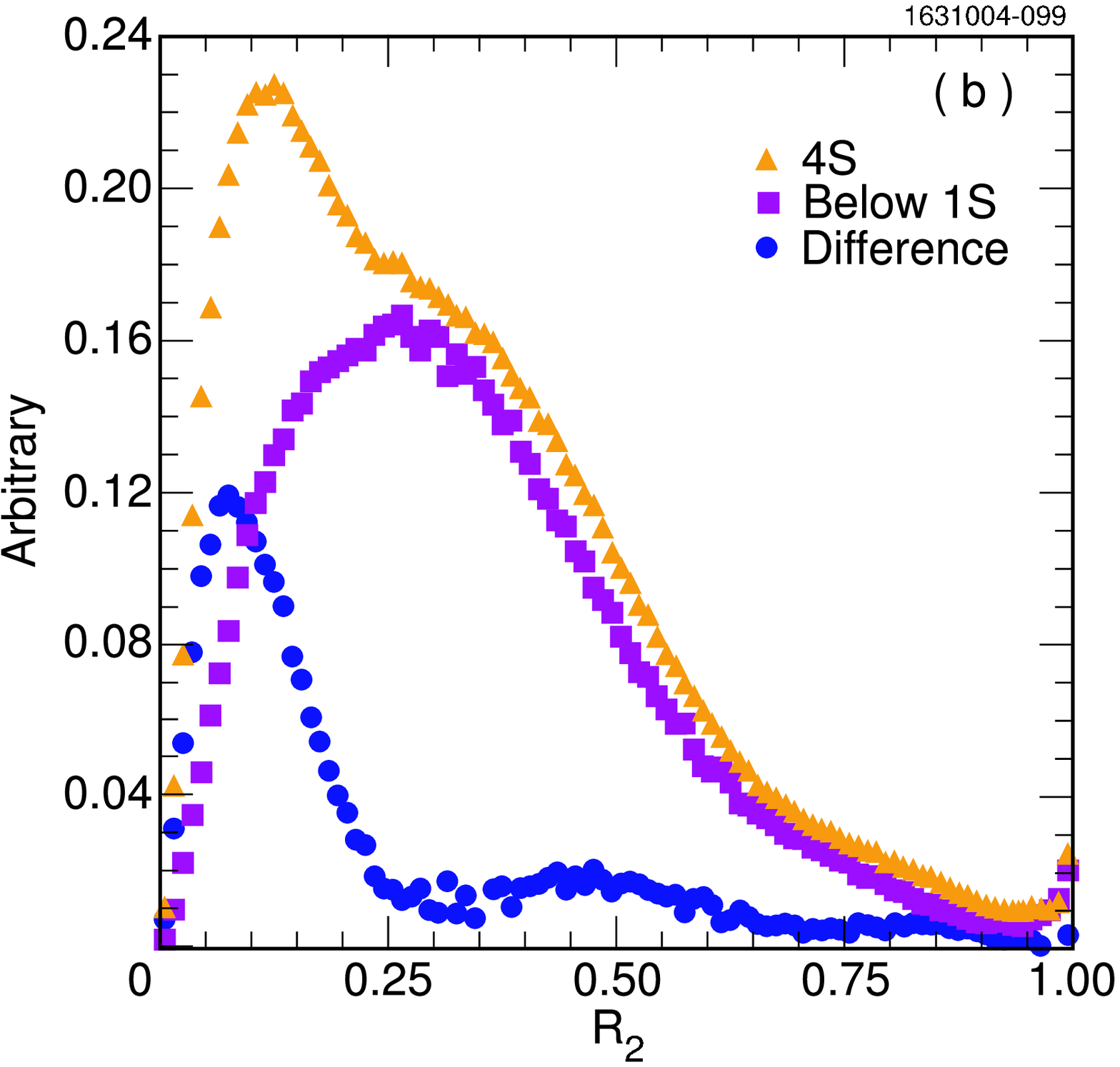,height=3in}
\caption{The $R_2$
distribution above $\Lambda_b$ threshold compared with
below-$\Upsilon(4S)$ data (a) and $\Upsilon(4S)$ on resonance data
compared with below-$\Upsilon(1S)$ data (b). Circles show the
subtracted distributions.} \label{fig:boost}
\end{figure}

To subtract four-flavor continuum background we use data taken at
a $E_{CM}$ 30 MeV below the $\Upsilon (4S)$ mass. Since we make a
specific cut on $R_2$ we need to take into account that the shape
of the $R_2$ distribution can change when the $E_{CM}$ changes. The
$R_2$ distribution from below-$\Upsilon(4S)$ data is compared with
the distribution using data taken in the $\Lambda_b$ scan region
in Fig.~\ref{fig:boost}(a). The data are normalized by luminosity
and $1/s$, where $s$ is the square of the center-of-mass energy.
The distributions differ in two respects. The first is the obvious
enhancement at small $R_2$ values in the $\Lambda_b$ scan region
giving evidence for $b\overline{b}$ production. The second is the
disagreement in shape at values of $R_2~>~0.5$, where
$b\overline{b}$ production is absent.

\begin{figure}[hbt]
\centerline{\epsfig{figure=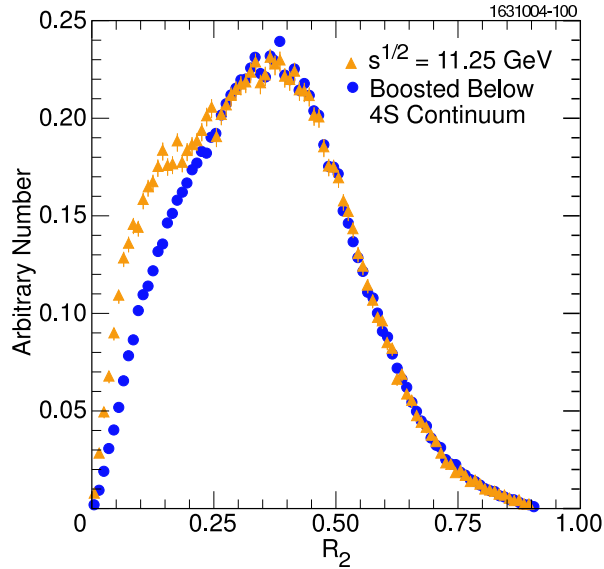,height=3in}}
        \caption{ $R_2$ Distribution at one energy point above $\Lambda_b$
        threshold compared with below-$\Upsilon(4S)$ after the boost (data).}
        \label{fig:boosted}
\end{figure}

We confirm this change in shape with energy by comparing
$\Upsilon(4S)$ ``on-resonance" data and below-$\Upsilon(1S)$
resonance data ($E_{CM}$=9.43 GeV) in Fig.~\ref{fig:boost}(b). The
subtracted spectra show an anomalous peak near $R_2 = 0.5 $. The
number of events in this peak can be as large as $\sim$30\% of the
total number of \bb events in higher $E_{CM}$ region. Thus, it is
important to transform correctly the below-$\Upsilon(4S)$
resonance data in order to correctly subtract the background when
we apply a tight $R_2$ requirement. Simple kinematic
considerations suggest that $R'_2(E')/R_2(E) \sim E'/E$, where $E'>E$.  The
boundary considerations that at $R_2$ values of both 0 and 1 
the initial and corrected distributions be equal, result
in a simple parameterization of the corrected, or ``boosted" $R_2$
distribution:
\begin{equation}
R'_2(E')= \frac{E'}{E} R_2(E)+\left( 1-\frac{E'}{E} \right)
R^2_2(E)~. \label{eq:boost}
\end{equation}
This expression describes the energy dependence of the $R_2$ shape
excellently. In Fig.~\ref{fig:boosted} we compare the boosted
$R_2$ distribution for below-$\Upsilon(4S)$ data, normalized by
luminosity and $1/s$, with the same distribution for the high
energy data. The distributions match above $R_2$ of 0.5, as
required.

We have several strategies for observing the production of \LLbar
events. One possibility is to look for enhancements in the (1) \bb
cross-section. Another is to look for an increase in (2) $\Lambda$ or (3)
anti-proton production. We don't use protons because there is a large
background rate from hadron interactions in the beam pipe and from
residual beam gas collisions. $\Lambda$'s are promising because we expect that $\Lambda_b^o\to\Lambda_c X$ has a large branching ratio, $\sim$96\% and $\Lambda_c^+\to \Lambda X$ is approximately 50\%.
Detecting anti-protons is very promising because
$\Lambda_b^0$ decays always produce either one proton or neutron.
In the case of non-resonant \LLbar production we can expect that
the cross-section will increase from zero at threshold to some
constant fraction of the total \bb cross-section. In order to ascertain an optimal search strategy, we assume this
fraction is 7.9\%, as predicted by the JETSET 7.3 Monte Carlo
model \cite{lambda_ratio}. This is consistent with the PDG value
for \bb $\to$ baryon of 10\% \cite{PDG}. Further support for this
value comes from the ratio of $\Lambda_c \overline{\Lambda}_c$ to
$c\overline{c}$ rates. As input to this estimate we use a measured
value of $\mathcal{B}(\Lambda_c^+ \to
pK^-\pi^+)\times\sigma(\Lambda_c^+) = (10.0\pm1.5\pm1.5)$ pb
\cite{cleo_lambdac}, from our below-$\Upsilon(4S)$ continuum data
sample. We take the $c\overline{c}$ cross section as 4/10 of the
total hadronic cross section, implying
$\sigma(c\overline{c})=1.12\pm0.02$ nb \cite{CLEOR}, and we use the
PDG mean value for $\mathcal{B}(\Lambda_c^+ \to pK^-\pi^+)=(5.0\pm
1.3)\%$ \cite{PDG}, yielding the ratio or
$\Lambda_c\overline{\Lambda_c}$/$c\overline{c} = (8.9\pm 3.0)$\%.

\begin{figure}[hbt]
    \begin{center}
    \epsfig{figure=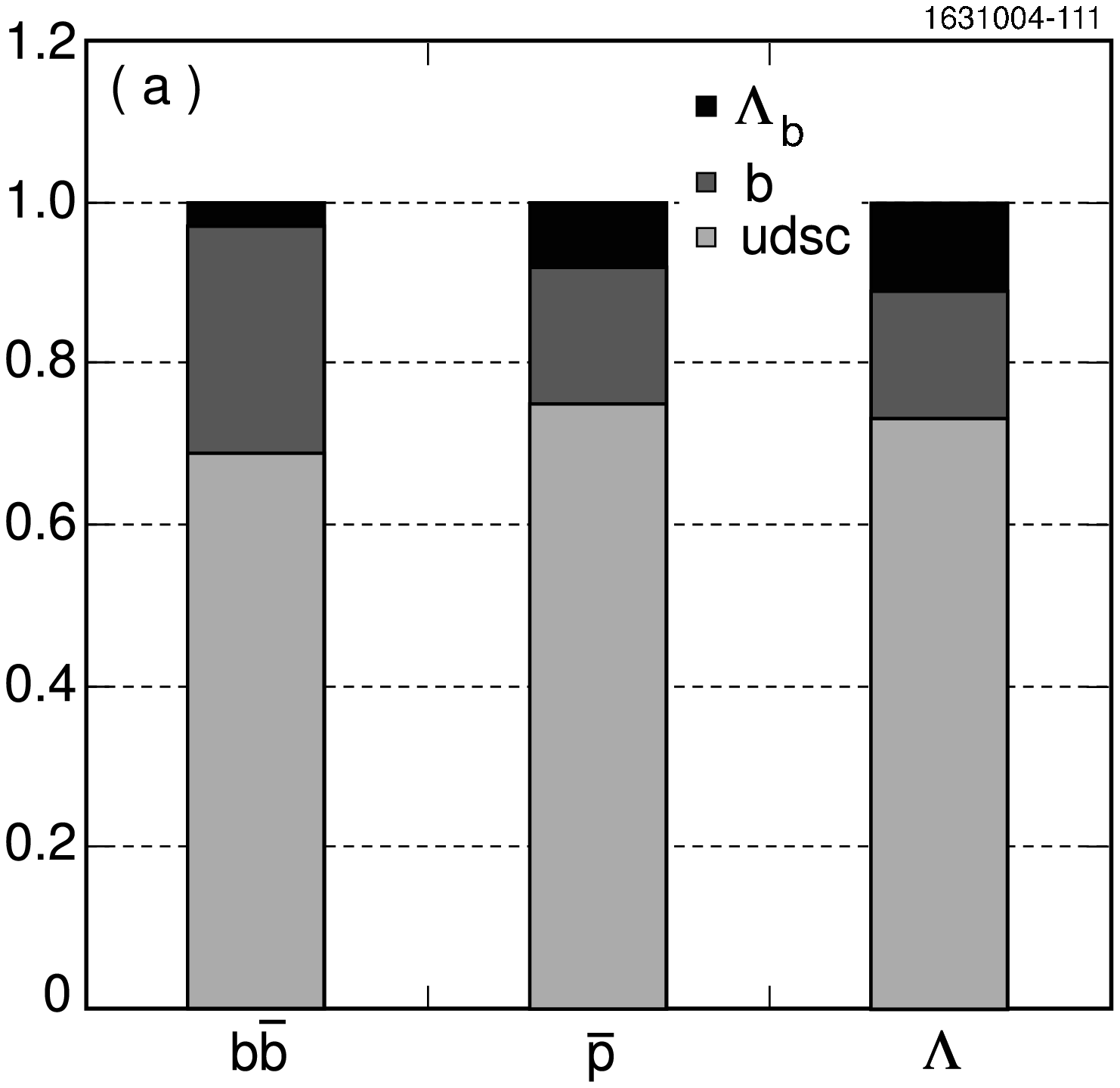,height=2.8in}
    \epsfig{figure=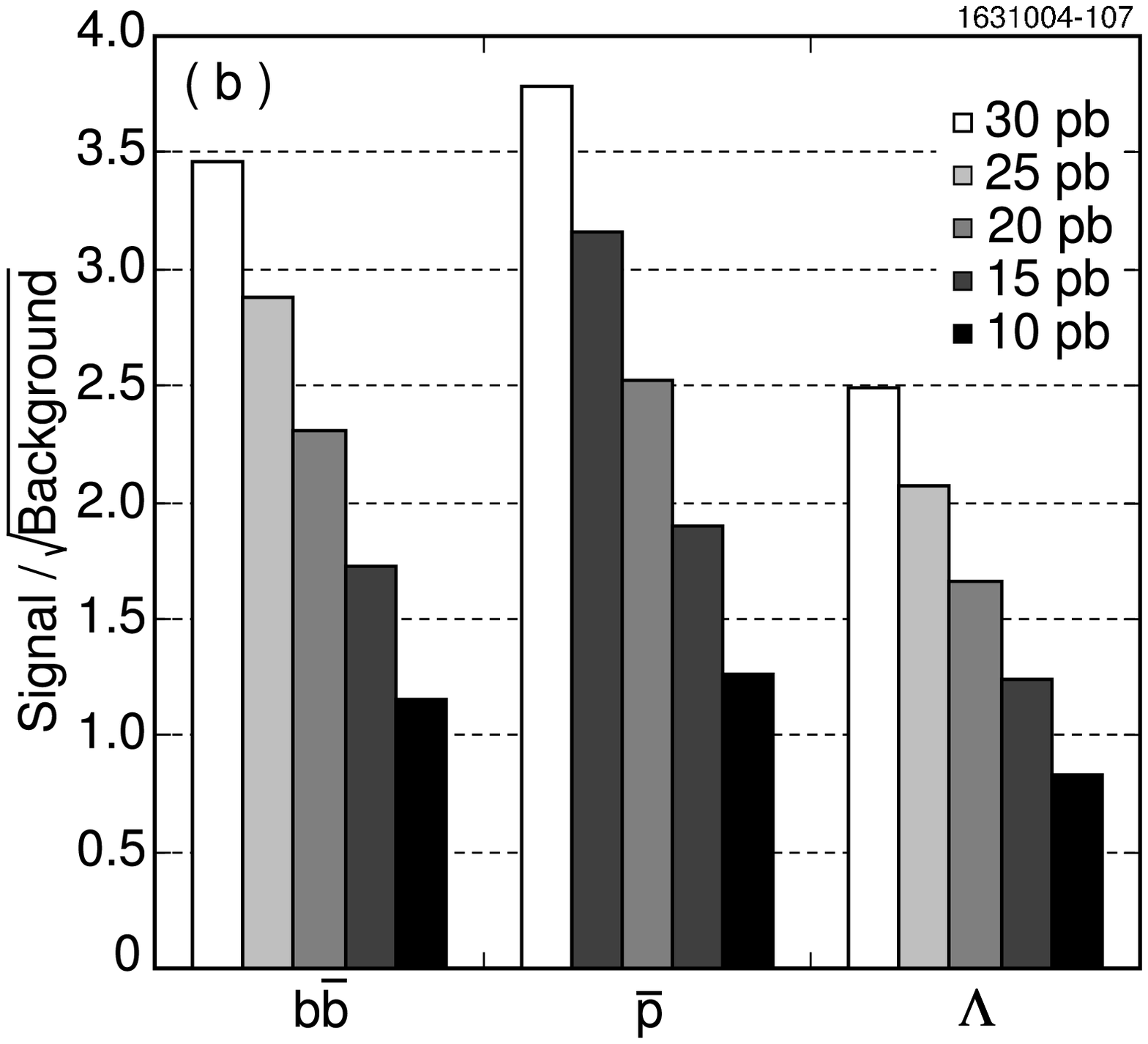,height=2.8in}
    \caption{(a) Relative yield of the $udsc$ (lower), $b$ (middle) and
    $\Lambda_b$ (upper) visible cross section for the inclusive selection
    of \bb, $\overline{p}$ and $\Lambda$ assuming a 7.9\% increase of the total \bb
    cross section above \LLbar threshold. (b) ${\rm Signal/\sqrt{Background}}$ for
    different analysis strategies and cross-sections.}
    \label{signoise}
    \end{center}
    \end{figure}

The relative size of the \LLbar component for our different search
strategies is shown in Fig.~\ref{signoise}(a). Here we normalized
the MC simulated five-flavor visible hadronic cross section to
unity, defined here as ``continuum'' $udsc$ and $b$, and then
added the signal \LLbar to the total $udscb$ cross section(i.e.,
the \LLbar enhancement here represents an additional 7.9\% above
expected inclusive \bb hadronic cross-section, rather than simply
presenting an additional channel available to \bb hadronization).
$\Lambda$'s have the highest relative yield closely followed by
anti-protons. We optimize our search criteria by maximizing signal
divided by square root of the background, $S/\sqrt{B}$, for our
different search methods. The results are summarized in
Fig.~\ref{signoise}(b), where we show the statistical significance
for signal we obtain for different analysis strategies for
different \LLbar cross-sections (statistical errors only).

Our studies indicate that baryon production (namely anti-protons
and $\Lambda$'s) is the most sensitive measure of \LLbar . However, the systematic
uncertainties in $\Lambda_b \to$ protons and $\Lambda_b \to
\Lambda$ diminish their sensitivity relative to inclusive
$b\overline{b}$ production. We also considered
identifying $\Lambda$'s and protons with an additional lepton in the event but these methods
offer less significance. The efficiencies for detecting hadronic
events, and more importantly, for detecting events with one or
more protons are listed in Table~\ref{tab:eff}; their evaluation will be
discussed in more detail in the next section.

We use both charged particle ionization loss in the drift chamber
(dE/dx) and RICH information to identify anti-protons. The RICH is used for momenta larger
than 1 GeV. Information on the angle of detected Cherenkov
photons is translated into a likelihood of a given photon being due
to a particular particle. Contributions from all photons
associated with a particular track are then summed to form an
overall likelihood denoted as ${\cal L}_i$ for each particle
hypothesis. To differentiate between kaon and proton candidates, we
use the difference: $-\log({\cal L}_{K})+\log({\cal L}_{proton}$).
This cut is set at -4. To
utilize the dE/dx information we calculate  $\sigma_{K}$ as the
difference between the expected ionization loss for a kaon and the
measured loss divided by the measurement error.  Similarly,
$\sigma_{proton}$ is defined  in the same manner using the expected
ionization for a proton.

We use both the RICH and dE/dx to select anti-proton candidates in the following manner: (a) If neither the RICH
nor dE/dx information is available, then the track is rejected. (b)
If dE/dx is available and RICH is not then we insist that proton
candidates have $PID_{dE}\equiv\sigma_{K}^2-\sigma_{proton}^2 <0$ (c) If RICH information is available
and dE/dx is not available, then we require that
$PID_{RICH}\equiv -\log({\cal L}_{K})+\log({\cal L}_{proton})<-4$. (d) If both dE/dx and RICH
information are available, we require that $(PID_{dE}+PID_{RICH})
<-4$.

$\Lambda$ candidates are formed from a pair of oppositely charged tracks one of which is consistent with a proton or anti-proton hypothesis, with a looser criteria than that stated above, which are constrained to
come from a single vertex. We also require that the invariant mass be within 5 times the width of the $\Lambda$ mass peak, which has an r.m.s. width of 1.4 MeV.

\subsection{Efficiency Determinations}
To derive event selection efficiencies we simulated hadronic events using
the JETSET 7.3 $q \overline{q}$ event generator \cite{Jetset},
then followed through the full GEANT 3.21-based \cite{cleog}
CLEO-III detector simulation. For five-flavor hadronic
and \LLbar events in the $\Lambda_b$ scan region, we generated
Monte Carlo samples using the same generator with the properties
described in section~\ref{sec:two}. The efficiencies obtained from these
simulations are presented in Table~\ref{tab:eff}, where we list the both the hadronic event selection efficiency and the efficiency for detecting a hadronic event with an anti-proton. These efficiencies include the branching ratios for the various processes into anti-protons in the second column. We take ${\cal{B}}(\Lambda_b^o\to \overline{p} X)=0.50$. The row for
$b\overline{b}$ includes only $B$ meson production with additional pions
allowed. As one would expect, the
efficiencies for $b\overline{b}$ and \LLbar are very similar. The slightly
lower efficiency for \LLbar arises from higher average jettiness
for \LLbar events.
\begin{table*}[htb]
\begin{center}
\caption{\label{tab:eff}
Selection efficiencies for hadronic events and
those with anti-protons.}
\vspace{0.2cm}
\begin{tabular}{lccr}
\hline\hline 
Data samples & Selection efficiency for & Selection efficiency for \\
             & hadronic events (\%) & hadronic events with an $\overline{p}$ (\%) \\ \hline
Below-$\Upsilon(4S)$ continuum & 25.5 $\pm$ 0.2 $\pm$ 0.8& 2.1 $\pm$ 0.1 $\pm$ 0.1\\
\LLbar & 85.5 $\pm$ 0.9 $\pm$ 2.6& 26.8 $\pm$ 0.1 $\pm$ 5.4\\
4 flavor ($udsc$) continuum & 21.9 $\pm$ 0.4 $\pm$ 0.7& 1.8 $\pm$ 0.2 $\pm$ 0.1\\
at $E_{beam} \sim m(\Lambda_b)$ \\
$b \overline{b}$ & 89.9 $\pm$ 1.2 $\pm$ 2.7& 4.0 $\pm$ 0.2 $\pm$ 0.3\\
5 flavor ($udscb$) continuum & 28.1 $\pm$ 2.5 $\pm$ 0.8& 2.0 $\pm$ 0.3 $\pm$ 0.2\\
$\tau \overline{\tau}$ & 0.024 $\pm$ 0.005 $\pm$ 0.001& $ < 10^{-5}$ \\
\hline\hline 
\end{tabular}
\end{center}
\end{table*}


The errors listed in Table~\ref{tab:eff} are statistical and systematic,
respectively. The systematic error for the hadronic event
selection requirement is estimated from the variation in the
number of hadronic events (corrected by efficiency and background)
when changing selection requirements. The systematic error for
the proton identification has been evaluated from proton
efficiency measurements using reconstructed $\Lambda$ events
from data and then comparing with the equivalent MC estimation.

Our simulations also give us the selection efficiency for detecting an event containing either a $\Lambda$ or an $\overline{\Lambda}$ from $\Lambda_b\overline{\Lambda}_b$ decay of {$16.6\pm 0.1_{-0.0}^{+1.0}$\%, including the ${\cal{B}}(\Lambda\to p\pi^-)$. 
Note that the 
PDG world average for ${\cal{B}}(\Lambda_c
\rightarrow p~ {\rm anything}$) is  ($50 \pm 16$)\%.
Similarly ${\cal{B}}(\Lambda_c \rightarrow \Lambda~ {\rm
anything}$) is ($35 \pm 11 $)\%~\cite{PDG}. The errors on these rates will be included separately as systematic effects.

\subsection{Systematic Errors}

The systematic errors in determining \LLbar production are given
in Table~\ref{tab:sys}. The largest error is due to the unknown branching
fraction of $\mathcal{B}(\Lambda_c \to p X)$ to which we assign a
32\% error. We also include errors on the hadron selection
efficiency and the background in the hadronic event sample,
evaluated by varying our selection criteria as well as taking into
account the variation with $E_{CM}$, the anti-proton
identification efficiency evaluated by examining a larger sample
of $\Lambda\to p\pi^-$ data, and the luminosity measurement
uncertainty estimated as $1\%$~ \cite{lumi_error}.

The total systematic error found by adding these elements in
quadrature is 2.7\%, 32\% and 31\% on the determination of \LLbar
production using $b\overline{b}$, anti-protons and $\Lambda$'s,
respectively.

\begin{table}[htb]
\begin{center}
\caption{\label{tab:sys}List of systematic errors in determining \LLbar production } \vspace{0.2cm} \vspace{0.2cm}
\begin{tabular}{lccr}
\hline\hline 
Source & Error (\%) \\ \hline
Hadron efficiency & $\pm$3 \\
$\Lambda_b^o\to\Lambda_c^+ X$ branching ratio & $\pm$4\\
Proton identification efficiency & $\pm$4 \\
$\Lambda_c^+\to p X$ branching fraction & $\pm$32 \\
$\Lambda_c^+\to\Lambda X$ branching fraction & $\pm$31 \\
Total background of hadronic events & $\pm$2 \\
Luminosity & $\pm1$ \\
\hline\hline 
\end{tabular}
\end{center}
\end{table}

\section{The Estimated $b \overline{b}$ Cross Section}

The hadronic cross section is generally expressed in terms of its
ratio $R$ to the point cross section  $e^+e^- \to \mu^+ \mu^-$. To
search for resonant or non-resonant production of \LLbar in $e^+e^-$ collisions
we measure the $b \overline{b}$ cross section over the energy
range of the scan. Theoretically, $R_{b \overline b}$ can be
expressed as follows:
\begin{equation}
R_{b \overline b}= R_{b \overline b}^0\left[1+\alpha_s/
\pi+C_2(\alpha_s/ \pi)^2+C_3(\alpha_s/ \pi)^3 \right],
\label{eq:Rbb}
\end{equation}
where $R_{b \overline b}^0=N_cq_b$. $N_c$ is the number of quark
colors, $q_b$ is the $b$ quark charge and $\alpha_s$ is the strong
coupling constant. The constants are $C_2=1.409$ and $C_3=-11.767$
~\cite{chet}.
In our energy regime, we expect a value for $R_{b \overline b}$ of 0.35.

To find the $b \overline{b}$ cross section we subtract the $R_2$
four-flavor continuum data distribution from the higher energy
data, correct for the efficiency of the $R_2$ cut and the hadronic
selection criteria and divide by
the relevant luminosity. We use a value of the cross-section for
$e^+e^-\to\mu^+\mu^-$ equal to 86.8 nb/$s$, where $s$ is the square of the center-of-mass energy in units of GeV. However, we do not make a precise measurement
of \bb cross section due to uncertainties in the correct scaling
factors of two-photon events and initial state radiation contributions in different
energy regions. Here we wish to measure any possible enhancement
above the \LLbar threshold. Our results are presented in
Fig.~\ref{bbXsec} (a).
\begin{figure}[hbt]
\epsfig{figure=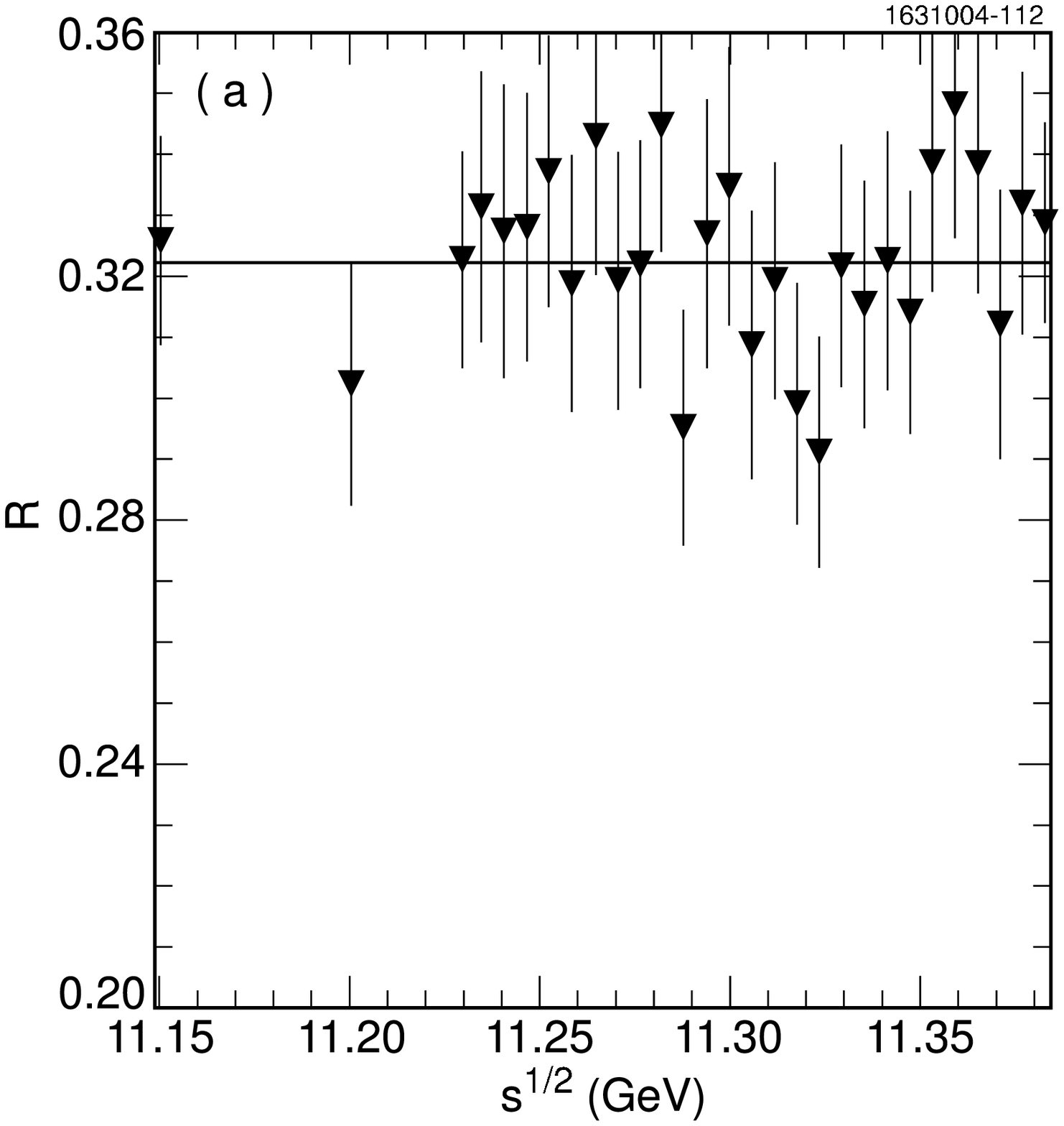,height=3in}
\epsfig{figure=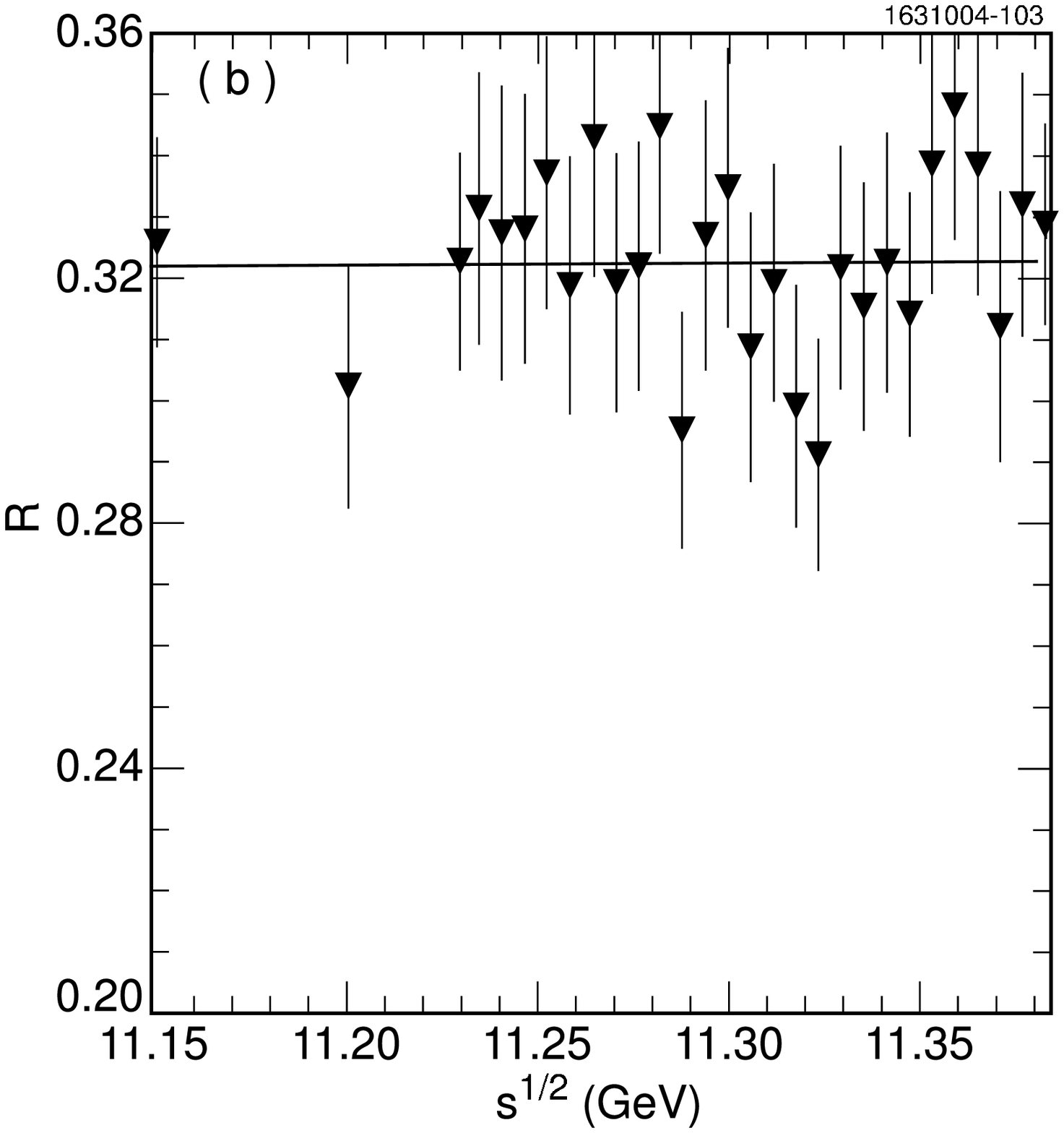,height=3in}
\caption{\label{bbXsec} The estimated $b \overline b$ cross
section in units of R. The error bars on the data points represent
both the statistical and the systematic errors summed in
quadrature. (a) The solid line shows a fit to a horizontal line.
(b) The solid line shows a fit to Equation~\ref{eq:bes}. The fits are described in
the text.}
\end{figure}

\section{Upper Limits on $\Lambda_b$ Production}

In this energy regime we expect that the R value will be constant
in the absence of any resonant or threshold increase due to \LLbar
production. There are no statistically significant excesses above
a constant value of R, suggesting no resonant production of
$b\overline{b}$ types of events. There is an important caveat
concerning the limit using the $b\overline{b}$ cross-section. It
may very well be that opening up the \LLbar channel comes at the
expense of a lower in rate of other channels so that the total
$b\overline{b}$ rate remains constant. Should this occur our
limit, in this (\bb) case, would be meaningless. In fact, a fit to
flat line for \bb yields a $\chi^2$ of 14.2 for 29 degrees of
freedom. This fit is shown on Fig.~\ref{bbXsec} (a).

We can look for an increase in \LLbar production that mimics the
threshold turn on as a function of center-of-mass energy of 
$e^+e^-\to \tau^+\tau^-$. The line in Fig.~\ref{bbXsec}
(b) represents a two-component fit. The first component is a
straight line without any slope allowed up to a $E_{CM}$ of 11.24 GeV,
twice the $\Lambda_b$ mass. The second component uses a shape
similar to one proposed by the BES collaboration~\cite{bes_paper}, but simplified
by explicitly calculating the Coulomb interaction and final state
radiation; the final form of this function is:
\begin{equation}
 \sigma(s)= A \times
\theta(\sqrt{s}-2m(\Lambda_b^0))(\sqrt{s}-2m(\Lambda_b^0))^{0.62}+R_0~~, \label{eq:bes}
\end{equation}
where $A$ is a fit parameter, $\theta(y)$ is step function, 0 for y$<$0
and 1 for y$>0$,
$m(\Lambda_b^0)$ is the mass and $R_0$ is the observed cross
section below threshold. (We are assuming this form applies only near threshold.)

\begin{figure}[hbt]
{\epsfig{figure=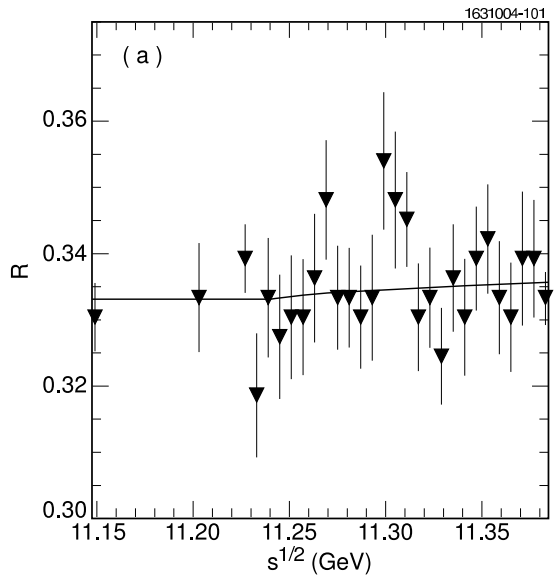,height=3in}}
\caption{\label{VisXsecp} The cross section for events with at
least one anti-proton normalized by $\sigma(e^+e^-\to \mu^+\mu^-)$. (The data have not been corrected for hadronic event efficiencies.) The solid lines show fits to
Equation~\ref{eq:bes}. The errors are statistical only.}
\end{figure}

The cross sections for events with anti-protons are shown in Fig.~\ref{VisXsecp}.
The data have been corrected for the momentum dependent efficiency of identifying anti-protons, but not for hadronic event selection.
The data are fit to the BES function given in
Eq.~(\ref{eq:bes}). The fitted parameters used to set upper limits
are listed in Table~\ref{tab:param}.

\begin{figure}[hbt]
{\epsfig{figure=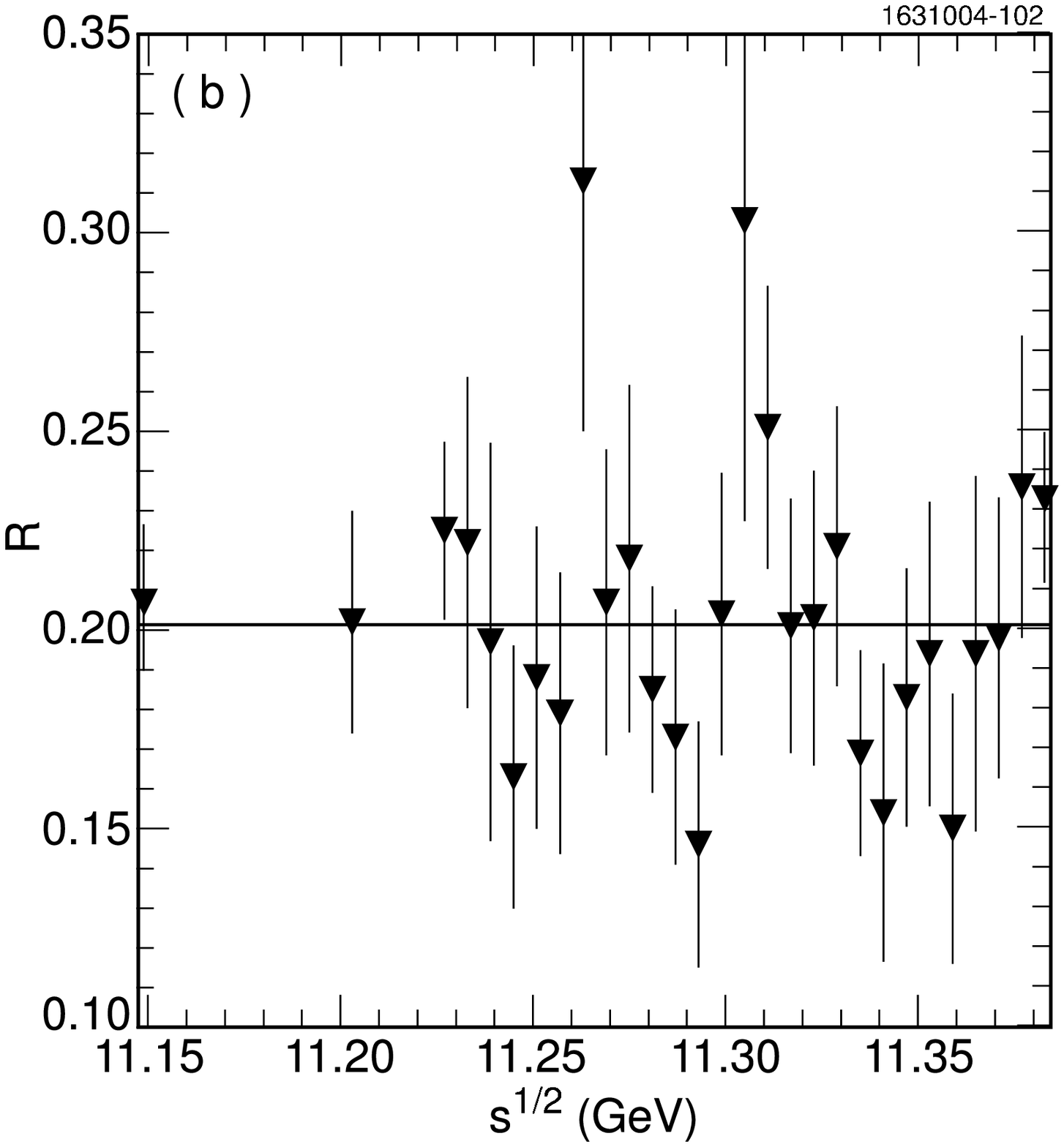,height=3in}}
\caption{\label{VisXsecL} The cross section for events with at
least one $\overline{\Lambda}$ normalized by $\sigma(e^+e^-\to \mu^+\mu^-)$.
(The data have not been corrected for hadronic event efficiencies.)The solid lines show fits to
Equation~\ref{eq:bes}. The errors are statistical only.}
\end{figure}

The cross sections for events with $\Lambda$'s are shown in Fig.~\ref{VisXsecL}.
The data have been corrected for both $\Lambda$ reconstruction efficiency and the branching ratio for \LLbar  into $\Lambda$ plus $\overline{\Lambda}$. 
The fit to data uses the BES function given in
Eq.~(\ref{eq:bes}). The fitted parameters used to set upper limits
are listed in Table~\ref{tab:param}.

\begin{table*}[htb]
\begin{center}
\caption{\label{tab:param}Numerical values of parameters found by fitting
Eq.~\ref{eq:bes} to our data.} \vspace{0.2cm} \vspace{0.2cm}
\begin{tabular}{lcc}
\hline\hline 
Selection criteria & $A_i$ & ${R_0}_i$ \\
\hline
\bb & $(0.21 \pm 3.82)\times 10^{-2}$ & 0.322 $\pm$ 0.007 \\
Anti-proton & $(0.84 \pm 1.20)\times 10^{-2}$ & 0.333 $\pm$ 0.002 \\
$\Lambda$ & $(0.15 \pm 5.49)\times 10^{-2}$ & 0.201 $\pm$ 0.010\\
\hline\hline 
Twice the $\Lambda_b$ mass is fixed to 11.24 GeV.

\end{tabular}
\end{center}
\end{table*}

There is no significant resonance peak in the scan range, nor any
evidence for a growth above threshold. 
Using these fits we calculate 95\% confidence level upper limits
for \LLbar production above threshold, as shown in
Fig.~\ref{upper_limit}. Here we take the upper limit as
\begin{equation}
\sigma(s)^{upper}_i = \left(A_i+1.64\times\delta A_i\right)
*\left(\sqrt{s}-2m(\Lambda_b^0)\right)^{0.62}/
\epsilon_i~~,
\end{equation}
where $A_i$ is the fit value from Table~\ref{tab:param}, $\delta A_i$ is its error, $\epsilon_i$ is the relative $\Lambda_b$ efficiency for each of the three
different methods of 0.95, 0.29, and
0.86, for $b\overline{b}$, $\overline{p}$ and $\Lambda$ searches,
respectively. The 0.95 results from the relative efficiency of continuum $b\overline{b}$ production to $\Lambda_b\overline{\Lambda}_b$, the 0.27 is the product of the $\Lambda_b\overline{\Lambda}_b$ decay rate into anti-protons and the efficiency of the hadronic event selection, and the 0.86 is hadronic event selection for $\Lambda_b\overline{\Lambda}_b$. The systematic errors are included only in the
limits using $b\overline{b}$ production. In the other two cases
the systematic errors on the inclusive $\overline{p}$ and
$\Lambda$ branching ratios worsen the upper limits by 32\% and
31\%, respectively.

We determine upper limits for production of a resonance that would
decay into \LLbar, similar in spirit to $\Upsilon$(4S) $\to
B\overline{B}$. Here we take two possible intervals for either a
narrow 6 MeV wide resonance or a wider, arbitrarily chosen, 18 MeV
resonance. For the first case we fit a horizontal line to our data
up to the \LLbar threshold of 11.24 GeV and then estimate the
upper limit for a cross-section excess in each 6 MeV interval of
center of mass energy. These 95\% confidence level upper limits
are shown in Fig.~\ref{res_upper_limit}(a).

For the second case, we fit all our data to a horizontal straight
line while excluding an 18 MeV wide interval of center-of-mass
energy. We then calculate the 95\% confidence level upper limit by
calculating the difference of the data relative to the fit line.
These limits are shown in Fig.~\ref{res_upper_limit}(b).

\begin{figure}[hbt]
\begin{center}
\epsfig{figure=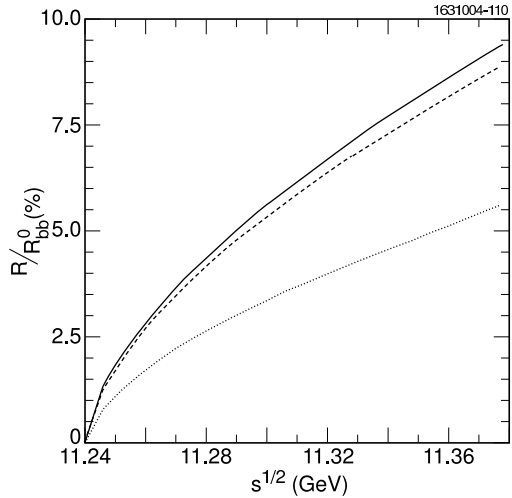,height=2.6in}\hspace{0.2in}

\caption{\label{upper_limit} The fractional upper limits at 95\% c. l. for
\LLbar production obtained using $\Lambda$ (solid line),
anti-proton (dashed line) and the \bb (dotted line) yields set by
using the BES function. For the \bb case only, systematic errors
have been included.}
\end{center}
\end{figure}

\begin{figure}[hbt]
\begin{center}
\epsfig{figure=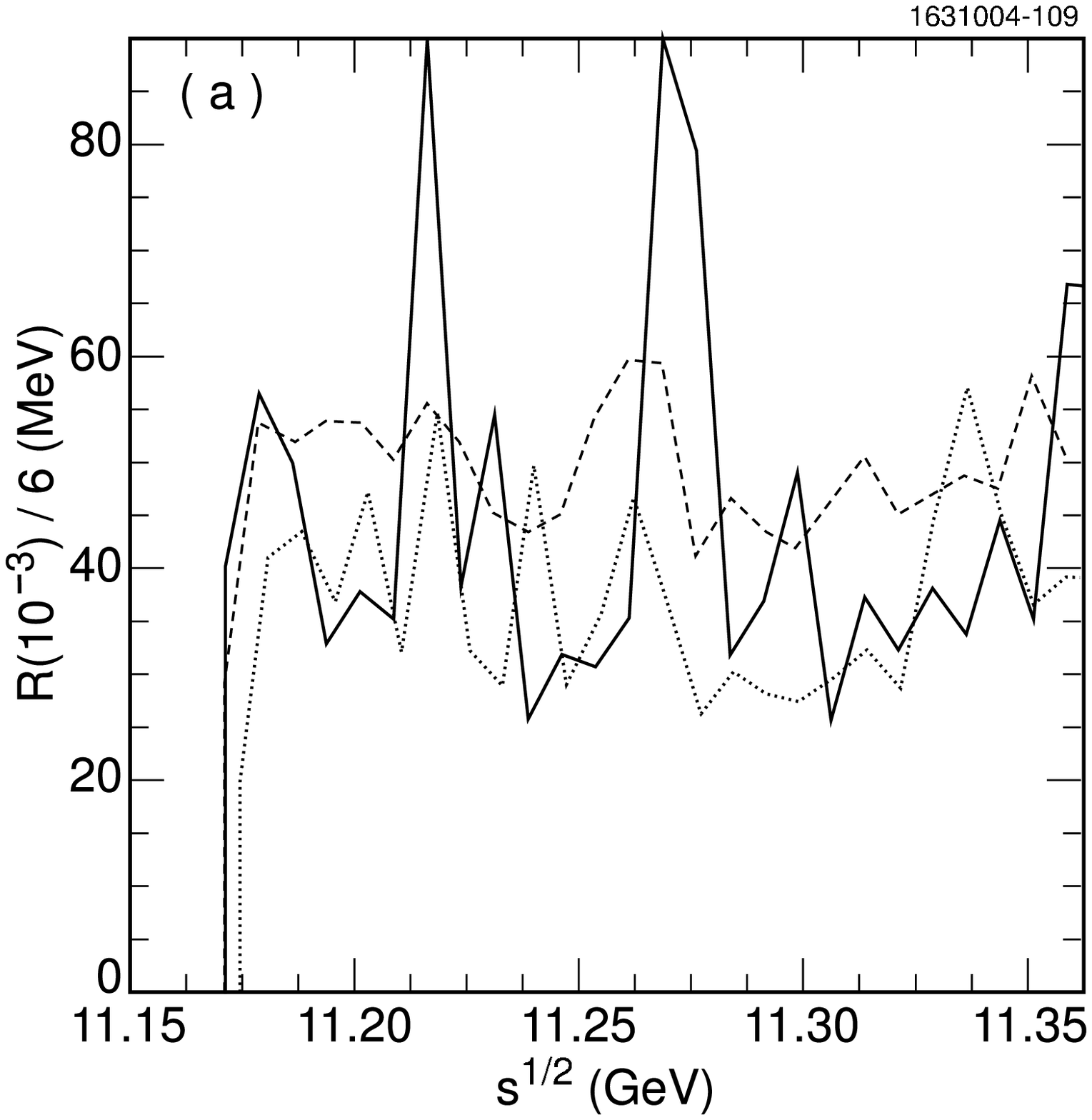,height=3in}
\epsfig{figure=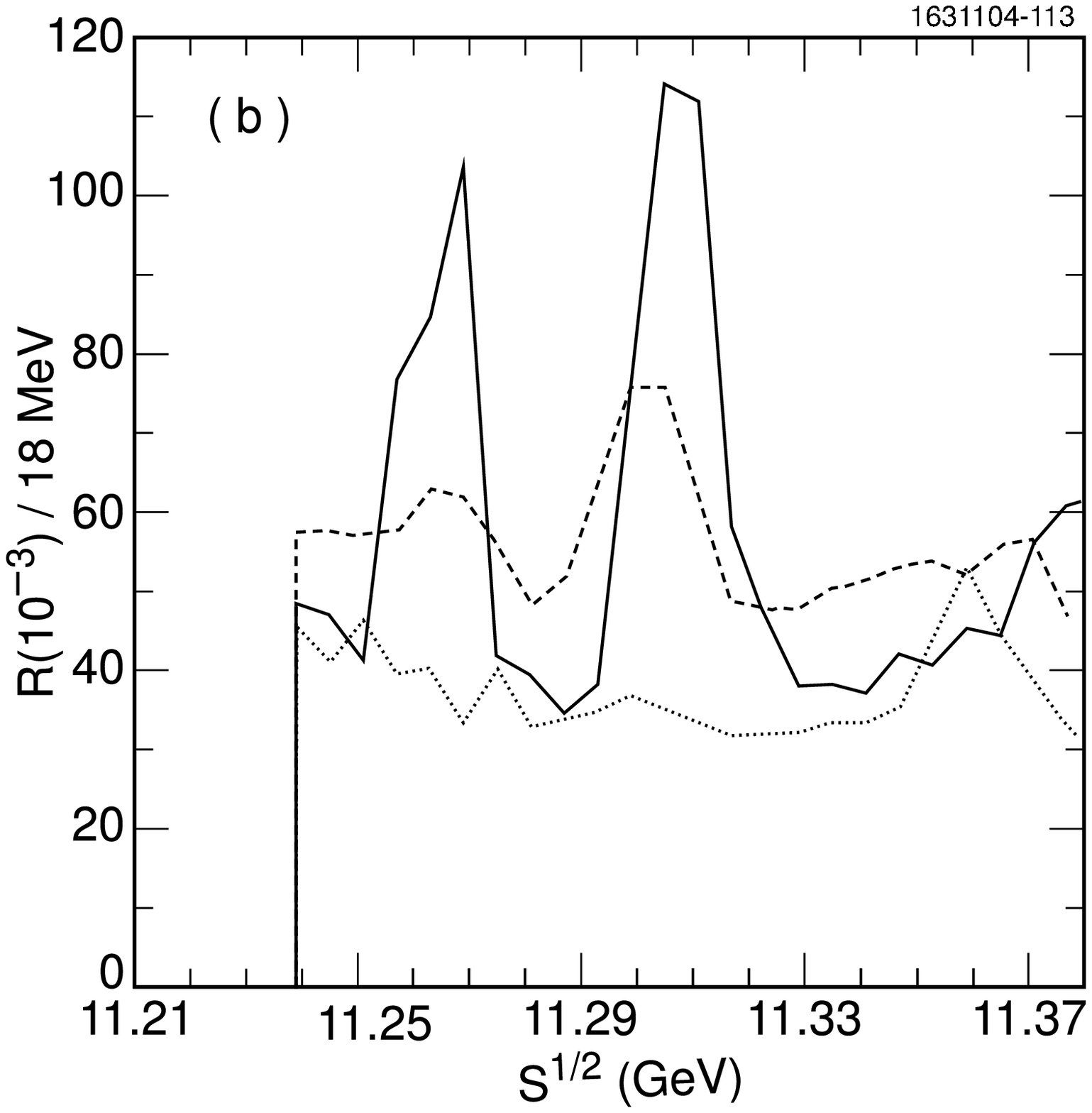,height=3in}
\caption{\label{res_upper_limit} Upper limits at 95\% c. l. for
\LLbar production obtained using $\Lambda$ (solid line),
anti-proton (dashed line) and the \bb (dotted line) yields. (a)
The upper limits have been set in 6 MeV center of mass energy
intervals in the scan region. (b) Upper limits in 18 MeV wide intervals. For
the \bb case only, systematic errors have been included.}
\end{center}
\end{figure}

No resonant enhancement reminiscent of the $\Upsilon (4S)$
resonance is observed. Using the threshold function we can set an
upper limit at our highest energy point of 11.383 GeV on the ratio
of $\Lambda_b^0$ to $b\overline{b}$ production. These limits are
given in Table~\ref{tab:ul}. For \bb production we use two values
- the first is $R_{b \overline b}^0$ as defined in
Eq.~\ref{eq:Rbb}; the second is  determined by fitting $R_{b
\overline b}$ values assuming no enhancement along scan range.
These values are $R_{b \overline b}^0 = 1/3$ and $R_{b \overline
b}= 0.322 \pm 0.004$.

The limits based on this function become lower toward lower energy
as we approach the production threshold. The anti-proton and
$\Lambda$ samples are somewhat correlated in that anti-protons
from $\overline{\Lambda}$ decay are often included in both
samples, so we choose not to combine these limits.

\begin{table*}[htb]
\begin{center}
\label{tab:ul} 
\caption{Upper limits at 95\% c.l. on the ratio of \LLbar to \bb
production at 11.383 GeV.} \vspace{0.2cm}\vspace{0.2cm}
\begin{tabular}{lcccc}
\hline\hline 
method & \multicolumn{2}{c}{95\% c.l. (statistical only)}

& \multicolumn{2}{c}{95\% c.l. (statistical \& systematic)}\\

& $R_{\Lambda_b^0\overline{\Lambda}_b^0}$/$R_{b\overline{b}}^0$ &
$R_{\Lambda_b^0\overline{\Lambda}_b^0}$/$R_{b\overline{b}}$ &
$R_{\Lambda_b^0\overline{\Lambda}_b^0}$/$R_{b\overline{b}}^0$ &
$R_{\Lambda_b^0\overline{\Lambda}_b^0}$/$R_{b\overline{b}}$\\
\cline{1-5}
\bb & - & - & 6.0\% & 6.2\%\\
Anti-proton & 9.2\% & 9.5\% & 12.2\% & 12.5\%\\
$\Lambda$ & 9.9\% & 10.2\% & 12.9\% & 13.3\%\\
\hline\hline 
\end{tabular}
\end{center}
\end{table*}

\section{Conclusions}

We do not observe any resonant or threshold enhancement of \LLbar
production in the center of mass energy region just above
threshold, resulting in 95\% confidence level upper limits on the
order of 0.05-0.10 units of R. The 95\% confidence level upper
limits from anti-proton and $\Lambda$ production are 12.8\% and
12.9\% of $R^0_{b\overline{b}}$, respectively, at our highest
energy point if they are modelled as a growth above threshold. In
order to effectively study $\Lambda_b$ decays at $e^+e^-$
machines, it may be necessary to go to higher center of mass
energies.

\end{document}